\documentclass[12pt]{article}  
\newcommand{\comma}{\;\; ,}
\newcommand{\period}{\;\; .}

\newcommand{\eq}{\; = \;}
\newcommand{\sep}{\;\; , \;\;}
\newcommand{\be}{\begin{equation}}
\newcommand{\bd}{\begin{displaymath}}
\newcommand{\ee}{\end{equation}}
\newcommand{\ed}{\end{displaymath}}
\newcommand{\ba}{\begin{eqnarray}}
\newcommand{\ea}{\end{eqnarray}}
\newcommand{\Wb}{\overline{W}}
\newcommand{\ab}{\overline{\alpha}}
\newcommand{\minus}{\! - \!}

\newcommand{\latticeL}{Figure 1}
\newcommand{\tqplane}{Figure 2}
\newcommand{\xyplane}{Figure 3}
\newcommand{\uqplane}{Figure 4}
{

\title{The ``inversion relation'' method for obtaining 
the free energy of the chiral Potts model}

\author{ R.J. Baxter\footnote{Fax: 61 2 6125 5549; e-mail: rj.baxter@apex.net.au.}
\\
{\protect \small Mathematical
Sciences Institute and Theoretical Physics, I.A.S. }\\
{\protect \small  Building 27, The Australian National University,
 Canberra, A.C.T. 0200, Australia\footnote{This work supported in part by 
the Australian Research Council.}  }
}

\begin{document}


\maketitle

}

\abstract{We derive the free energy of the chiral Potts model by the infinite lattice 
``inversion relation'' method. This method is non-rigorous in that it always needs 
appropriate analyticity assumptions. Guided by 
previous calculations based on exact finite-lattice functional relations, we find 
that in addition to the usual assumption that the free energy be analytic and bounded in some
principal domain of the rapidity parameter space that includes the physical regime, 
we also need a much less obvious symmetry.
We can then obtain the free energy by Wiener-Hopf ~factorization in the complex planes of 
appropriate variables. Together with the inversion relation, this symmetry 
relates the values of the free energy in all neighbouring domains to those
in the principal domain.


}

\section{Introduction}

Onsager calculated the free energy of the two-dimensional Ising model by 
setting up an 
algebra that contained the row-to-row transfer matrix \cite{Onsager44}.
Kaufman simplified the derivation by using spinor operators 
(i.e. a Clifford algebra)\cite{Kaufman49}, 
and Kasteleyn  showed that the result could be obtained quite 
directly by pfaffians \cite{Kasteleyn63}.

Most other solvable models do not appear to be amenable to such algebraic (in particular  
free-fermionic) methods. For these there are three main methods of approach (excluding  
the Bethe ansatz eigenvector method): the first and third 
depend explicitly on the Yang-Baxter or star-triangle relation, the second implicitly.

The first is to set up functional relations for the transfer matrix $T$. These 
define all the eigenvalues of $T$ (for a finite system), so in 
principle give the exact free energy for a lattice of $M$ rows and $L$ columns. Usually 
one can solve these explicitly only in the large-lattice limit.

The second is the inversion relation method, due to Stroganov \cite{Stroganov79}. In this one 
writes down 
rotation and inversion relations satisfied by the free energy of the infinite 
system. They are functional relations (sometimes called difference equations), 
wherein the free energy is regarded as a function of the rapidity variables
(or of their difference, which is the  spectral variable). These relations do 
not by themselves define  the free energy, any more than the relation $f(z+1) = 2 - f(z)$
defines a function $f(z)$ in the complex plane. It is
essential to make appropriate analyticity assumptions to complete the relations: for 
instance 
in our simple example, if we know that $f(z)$ is analytic and bounded in the vertical strip
$0 \leq \Re (z) \leq 1$, then it must be analytic and bounded in the whole plane. 
By Liouville's theorem it is therefore  a constant, so $f(z) = 1$.

The third method depends on using the star-triangle relation to relate the free 
energies of the triangular and honeycomb lattices, differentiating this and using the 
fact that the derivatives are local correlations that depend only on two rapidities, 
rather than three. This gives a partial differential equation for the derivatives. As 
far as the  author is aware,  this method has been applied only to the Ising and 
chiral Potts models \cite{BaxterEnting78, Baxter88}. 

The free energy of the chiral Potts model  has been 
obtained only by the first and third methods. The result of the first 
\cite{Seoul90, Baxter90} is an explicit  double integral, 
while the second \cite{Baxter88} gives a hierarchy of equations that 
implicitly define the  free energy (and can be used to obtain the critical 
exponent $\alpha$). Only recently  \cite{Baxter00} has 
it been verifed that the two results are equivalent.

The object of this paper is to use the second method - the inversion relation method - to 
obtain the free energy of the $N$-state chiral Potts model in the infinite-lattice limit. We 
do this by starting from the functional relations of method 1, taking the 
large-lattice limit, and showing that 

a) the resulting equations can be derived from the inversion and rotation relations,

b) they can be solved by making certain analyticity assumptions. In fact we know 
these assumptions are valid from method 1, but we offer plausibility arguments in 
their favour that are independent of method 1.

We then use Wiener-Hopf factorization methods to  solve these infinite-lattice 
functional relations, and inevitably 
obtain the double integral result of method 1. 

In later papers we intend to  further investigate the analyticity properties of the 
free energy, or rather 
of the partition-function per site $\kappa$, as a function of the two rapidity variables $p$ and $q$. 
 It is a meromorphic function on
a Riemann surface of infinite genus. Each sheet of this surface can be associated with a point
on a lattice in $2N-1$-dimensional space (for $N > 2$). If one fixes one of the two rapidities 
$p$ or $q$, then
the remaining single-rapidity space is only $N$-dimensional.
(The fact that the two-rapidity surface is only $2N-1$-dimensional, rather than $2N$, is a 
manifestation of a weak ``rapidity difference property''. Unlike other planar models, it is not 
obvious that this property is particularly helpful.)

This surface is {\em not}
 (for $N > 2$) contained in the
surface generated by the hyperelliptic function variables known to be associated with the 
chiral Potts model
\cite{Baxter91}. Thus it is not a single valued-function of these particular variables, so 
cannot be immediately expressed in terms of hyperelliptic functions of them. However,
the surface for the associated function $\tau_2(t_q)$ is associated with a space of 
one dimension less, and  {\em is} the space spanned by the hyperelliptic variables (after a 
simple transformation), so hyperelliptic 
functions may still be useful in this problem.

There are two reasons for this investigation. One is that a large amount of 
fascinating mathematics has grown out of solvable models, much of it connected with
the explicit Jacobi elliptic function parametrization that is used in simpler
models such as hard hexagons  \cite{Baxter81}. Does this generalize to the 
chiral Potts model and its associated hyperelliptic function parametrization?

The other reason is the continuing problem of the intriguing conjecture for the 
spontaneous magnetizations of the chiral Potts model \cite{Howes83,Albertini89,
HenkelLacki89}. This remains unproven,
but there is an infinite-lattice functional relation for it 
\cite{Baxter98a,Baxter98b,Baxter98c} which is 
much like the inversion relation for the free energy. If one knows how to  solve 
the latter, then one might hope to use similar
techniques to solve the former, and hence verify the conjecture. Only time will tell!

One interesting  result in this paper is (\ref{2ndratio}):
an unexpected simple symmetry relation satisfied by the free energy of 
the chiral Potts model.

\section{Functional relations for the transfer matrices}

Consider the square lattice $\cal L$, drawn diagonally as in  \latticeL, with $2 M$ rows of  
$L$ sites  and
periodic boundary conditions. At each site $i$ there is a ``spin'' $\sigma_i$, which takes 
the values $0, \ldots , N-1$. Adjacent spins $i$ and $j$ (with $i$ below $j$) interact with 
Boltzmann weight $W_{pq}( \sigma_i - \sigma_j )$ for SW $\rightarrow$ NE edges, 
$\Wb_{pq}( \sigma_i - \sigma_j )$ for SE $\rightarrow$ NW edges. The 
parameters $p, q$ will now be defined.

\begin{figure}[hbt]
\begin{picture}(420,300) (-40,-100)
\multiput(0,0)(80,0){5}{\circle*{5}}
\multiput(0,80)(80,0){5}{\circle*{5}}
\multiput(0,160)(80,0){5}{\circle*{5}}
\multiput(40,40)(80,0){4}{\circle*{5}}
\multiput(40,120)(80,0){4}{\circle*{5}}
\multiput(-4,100)(5,0){69}{.}
\multiput(-4,140)(5,0){69}{.}
\multiput(178,-18)(0,5){36}{.}
\multiput(218,-18)(0,5){36}{.}
\multiput(258,-18)(0,5){36}{.}
\put(0,0) {\line (1,1) {160}}
\put(160,0) {\line (-1,1) {160}}
\put(160,0) {\line (1,1) {160}}
\put(320,0) {\line (-1,1) {160}}
\put(80,0) {\line (-1,1) {80}}
\put(0,80) {\line (1,1) {80}}
\put(80,0) {\line (1,1) {160}}
\put(240,0) {\line (-1,1) {160}}
\put(240,0) {\line (1,1) {80}}
\put(320,80) {\line (-1,1) {80}}
\thicklines
\put(184,104) {\vector (1,1) {6}}
\put(216,104) {\vector (-1,1) {6}}
\thinlines
\put(222,107) {$\overline{W}_{\! p'q}$}
\put(182,89) {$W_{\! pq}$}
\put(343,98) {$q$}
\put(343,138) {$q$}
\put(178,-30) {$p$}
\put(218,-30) {$p'$}
\put(258,-30) {$p$}
\put(-5,66) {$\sigma''_1$}
\put(75,66) {$\sigma''_2$} \put(-1,-14) {$\sigma_1$}
 \put(36,26) {$\sigma'_1$}
\put(79,-14) {$\sigma_2$} \put(116,26) {$\sigma'_2$}
 \put(239,-14){$\sigma_L$} \put(276,26) {$\sigma'_L$} 
\put(314,-14) {$\sigma_{L+1}$}\put(314,66) {$\sigma''_{L+1}$}
\put(159,66) {$i$} \put(239,66) {} 
\put(199,128) {$j$} 
\put(343,12) {$T_q$} 
\put(343,48) {$\hat{T}_q$} 
\put(0,-60) {Figure 1: The square lattice ${\cal L}$ of $2M$ (= 4)
rows with $L$ sites per row. }
\put(15,-75) {$T_q$ is the 
transfer matrix of  an odd row, $\hat{T}_q$ of an even row.
Three }
\put(15,-90) {vertical and two horizontal dotted rapidity lines
are shown.} 
\end{picture}
\end{figure}

Let $\omega = e^{2 \pi i /N}$ be the primitive $N$th root of unity, and
take $k, k'$ to be two  positive real constants, with $k,k' < 1$ such that
$ k^2 + {k'}^2 = 1$. 
Also, let $p = \{ x_p, y_p, \lambda_p, t_p , \mu_p\}$ be a set of complex parameters (``$p$-variables'')
related by
\be \label{relxy}
 k x_p^N = 1 - k'/\lambda_p \sep
 k y_p^N = 1 - k'\lambda_p \comma \ee
\be \label{deftlambda}
 x_p^N + y_p^N = k (1+ x_p^N y_p^N) \sep t_p = x_p y_p \sep  \lambda_p = \mu_p^N \period \ee

Only one of these variables is independent. In terms of the $a_p, b_p, c_p, d_p$ of ref. 
\cite{BPAY88}, $x_p = a_p/d_p$, $y_p = b_p/c_p$, $\mu_p = d_p/ c_p$. We can regard $p$ as a 
point on an algebraic curve in $(x_p, y_p, \lambda_p, t_p, \mu_p)$-space, and refer to it 
as a ``rapidity''.

Similarly, define  ``$q$-variables'' $q = \{ x_q, y_q, \lambda_q, t_q, \mu_q \} $. Then
the Boltzmann weight functions are
\ba \label{wts}
W_{pq}(n) \eq W_{pq}(n+N) &  = & (\mu_p/\mu_q)^n \prod_{j=1}^n (y_q - \omega^j x_p) /
(y_p - \omega^j x_q) \comma \nonumber \\
\Wb_{pq}(n) \eq \Wb_{pq}(n+N) &  = & (\mu_p\mu_q)^n \prod_{j=1}^n (\omega x_p - \omega^j x_q) /
(y_q - \omega^j y_p) \period  \ea

Note that $W_{pq}(0) = \Wb_{pq}(0) = 1$. We use this normalization throughout this paper. 
If $x_p, x_q, y_p, y_q, \omega x_p$ all lie on the unit circle and are ordered
anti-cyclically round it, then the Boltzmann weights $W_{pq}(n), \Wb_{pq}(n) $ are 
real and positive for all integers $n$. We refer to this case as the {\em physical regime}.

If $Z$ is the usual partition function, in the physical regime  we expect the limit
\be \label{pfnpsite}
\kappa_{pq} \eq \lim_{{\cal N} \rightarrow \infty}  Z^{1/{\cal N}} \ee
to exist, where $\cal N$ is the number of sites of the lattice, and the limit is 
to be taken so that the lattice is infinitely large in all directions. We refer to 
$\kappa_{pq}$ as the partition function per site. The free energy 
per site is
\be \label{freeenergy}
F_{pq} = -k_B {\cal T} \log \kappa_{pq} \comma \ee
$k_B$ being Boltzmann's constant and $\cal T$ the temperature.
Having taken this limit in the physical regime, we then define $\kappa_{pq}$ outside 
the regime by analytic 
continuation. We shall find that the resulting function $\kappa_{pq}$ lives on an infinite
Reimann surface.

Each edge of the lattice can be regarded as  associated with two rapidity lines, one 
vertical and one horizontal, as in  \latticeL. The rapidities may be different 
for different lines. In \ref{wts}, $p$ is the rapidity of the vertical line through 
the edge being considered, $q$ is the rapidity of the horizontal line.

As in \cite{BBP90}, we distinguish alternate columns of the lattice, assigning them
vertical rapidities $p$ and $p'$ as in  \latticeL. Then, as in \cite{BBP90}, 
we can define two $N^L$ by $N^L$ 
row-to-row transfer matrices $T$, $\hat{T}$, corresponding to the two types of 
row of the lattice. They have elements
\bd
T_{\sigma, \sigma '} \eq \prod_{j=1}^L W_{pq} (\sigma_j - {\sigma_j}')
\Wb_{p' q} (\sigma_{j+1} - {\sigma_j}') \comma \ed
\be \label{defT}
\hat{T}_{\sigma, \sigma '} \eq \prod_{j=1}^L \Wb_{pq} (\sigma_j - {\sigma_j}')
W_{p' q} (\sigma_{j} - {\sigma_{j+1}}') \period \ee

If there are $M$ rows of each type, then the partition function is
\be \label{partfn}
Z \eq {\rm Trace} (T \hat{T} )^M \period \ee

We regard $p$ and $p'$ as given, and view $T$, $\hat{T}$ as functions of 
the horizontal rapidity $q$. The parameter $\mu_q$ enters the rhs of (\ref{defT}) only 
via its $N$th power,
so they are uniqely defined if both $x_q$ and $y_q$ are prescribed. We write them as
$T(x_q,y_q)$, $\hat{T}(x_q,y_q)$.

Some scalar functions of $q$ that we shall need are
\be \label{defz}
z(t_q) \eq \left[ \omega \mu_p \mu_{p'} (t_p - t_q) (t_{p'} - t_q )/(y_p y_{p'})^2
\right]^L \comma \ee
\bd
z_j(t_q) \eq z(t_q) z(\omega t_q) \cdots z(\omega^{j-1} t_q ) \comma \ed
\bd \alpha_q \eq \alpha (\lambda_q) \eq  
\left[ \frac{k' (1- \lambda_p \lambda_q) (1- \lambda_{p'} \lambda_q) }
{k^2 \lambda_q y_p^N y_{p'}^N } \right]^L \eq
\left[ \frac{ (y_p^N- x_q^N ) (t_{p'}^N- t_q^N) }
{y_p^N y_{p'}^N  (x_{p'}^N- x_q^N) } \right]^L\comma \ed
\be \label{defalpha}
\ab_q \eq \alpha(1/\lambda_q) \sep \alpha_q \ab_q = z_N(t_q) \comma \ee
\be \label{defrpq}
r_{p',q}  = r_{p'}(x_q,y_q) =  \left\{ \frac{N (x_{p'} - x_q) (y_{p'} - y_q ) (t_{p'}^N - t_q^N) }
{(x_{p'}^N - x_q^N) (y_{p'}^N - y_q^N ) (t_{p'} - t_q) } \right\}^L \comma \ee
\be \label{hdef}
h_{j,q} \eq \prod_{m=1}^{j-1} \left\{ \frac{y_p y_{p'} (x_{p'} - \omega^m x_q)}
{(y_p - \omega^m x_q )(t_{p'}- \omega^m t_q ) } \right\}^L \period \ee

We shall also need the $N^L$ by $N^L$ cyclic shift matrix $X$, with elements
\be
X_{\sigma, \sigma'} \eq \prod_{j=1}^L \delta_{\sigma_j, \sigma'_j +1 } \period \ee

\subsubsection*{The $T \hat{T}$ relations} 
Without loss of generality, we can  take the integers $k, l$ of \cite{BBP90} to be
$0, j$. Then the functional relations (3.46) 
therein become
\be \label{3.46}
T(x_q, y_q) \, \hat{T} (y_q, \omega^j x_q) \eq r_{p',q} \, h_{j,q} \left[ \tau_j (t_q) +
z_j (t_q) \, X^j \, \tau_{N-j} (\omega^j t_q ) / \alpha_q \right] \comma \ee
for $j = 0, \ldots ,N$.

 Here $\tau_j(t_q)$ is the transfer matrix of the associated $\tau_j$ model defined in
eqns. (3.44) - (3.48) of \cite{BBP90} (which is related to the superintegrable chiral Potts 
model and to the model whose column transfer matrix is the $Q$ matrix of the six-vertex model
\cite{BazStrog90}). This matrix depends on $q$ only via the parameter $t_q$: in fact 
it is is  a polynomial in $t_q$ of degree $(j-1) L$.

\subsubsection*{The $\tau_2 T$ relation}
Similarly, equation (4.20) of \cite{BBP90} becomes
\bd  \tau_2 (t_q) \, T(\omega x_q, y_q)  = \left[ \frac{
 (y_p - \omega x_q ) (t_{p'} - t_q)}{y_p y_{p'} (x_{p'} - x_q) } \right]^L 
 \, T(x_q, y_q) \; + 
 \ed
\be \label{4.20}
\left[ \frac{ \omega \mu_p \mu_{p'} (x_{p'} - \omega x_q ) (t_{p} - \omega t_q)}{y_p y_{p'} 
(y_{p} - \omega^2 x_q) } \right]^L 
X \, T(\omega^2 x_q, y_q) \period \ee

Two other relations can immediately  be obtained from (\ref{3.46}) and (\ref{4.20}) by 
interchanging 
$p$ with $p'$ and $T$ with $\hat{T}$, while leaving $\tau_j(t_q)$ unchanged. As is 
shown in (4.22) - (4.30) of \cite{BBP90}, one can then  deduce from them two sets of relations involving 
only the functions $\tau_j(t_q)$, which we now give.

\subsubsection*{The $\tau_j$ relations}
The relations (4.27) of \cite{BBP90} are
\be \label{tau1}
\tau_j(t_q) \tau_2(\omega^{j-1} t_q) \eq z(\omega^{j-1} t_q ) X 
\tau_{j-1} (t_q) + \tau_{j+1} (t_q) \comma \ee
\be \label{tau2}
\tau_j(\omega t_q) \tau_2( t_q) \eq z(\omega t_q ) X 
\tau_{j-1} (\omega^2 t_q) + \tau_{j+1} (t_q) \comma \ee
for $j = 1 , \ldots , N$, where $\tau_{N+1}(t_q)$ is defined by
\be \label{tau3}
\tau_{N+1} (t_q) \eq z(t_q) X \tau_{N-1} (\omega t_q )  + (\alpha_q + \ab_q) \tau_1 (t_q) 
 \ee
and \be \label{tau01}
\tau_0(t_q) = 0 \sep \tau_1 (t_q) = 1 \comma \ee
i.e. $\tau_1 (t_q)$ is the identity matrix.

\subsubsection*{Invariances}

Apart from the relations $k^2+{k'}^2 = 1$, $ x_p^N + y_p^N = k (1+ x_p^N y_p^N)$, all the 
above equations are unchanged by
multiplying $x_p, x_q$, $y_p, y_q, t_p, t_q, k$ by $\epsilon, \epsilon, \epsilon, \epsilon, 
\epsilon^2, \epsilon^2$, $\epsilon^{-N}$, while leaving $\lambda_p ,\lambda_q , \mu_p, \mu_q$ 
and the functions $T(x_q,y_q), \tau_j(t_q), S(\lambda_q) $  unchanged. This invariance persists if 
in section 3 we write $\eta$ and $1/\eta$ as  $[(1-k')/k]^{2/N}$ and  $[(1+k')/k]^{2/N}$, respectively.  

Also, $z_j(t_q), \alpha_q, \ab_q,  h_{j,q}, \tau_j(t_q), \xi_q, S(\lambda_q)$ explicitly
contain the factors $\beta^{-2j}$, 
$\beta^{-N}$, $\beta^{-N}, \beta^{j-1}$, $\beta^{1-j}$, $\beta^{-N(N-1)/2}$, $\beta^{-N(N-1)/2}$,
where $\beta = (y_p y_{p'})^L$. These factors
are ``constants'' (independent of $q$) and cancel out of the functional relations
(\ref{3.46}) - (\ref{tau3}), so we could have re-defined the functions so as to remove them. 
We prefer to leave them in so as to make the previous invariance more explicit.

\subsubsection*{Comments}

Because of the star-triangle relation, the matrices $T, \hat{T}$ satisfy 
the commutation relations (2.32)- (2.33) of \cite{BBP90}. From this it follows that
there exist invertible matrices $P_1$,$P_2$, independent of the horizontal rapidity $q$, 
such that $P_1^{-1}  T(x_q,y_q) P_2$, $P_2^{-1} \hat{T}(x_q,y_q) P_1$,
$P_1^{-1}  \tau_j(t_q) P_1$ are all diagonal matrices, for all $q$. In this sense the 
functional relations above can all be simultaneously diagonalized. Their diagonal
elements are then just scalar functional relations for each eigenvalue.

Hereinafter we shall we shall work in this diagonal representation. Further, we shall 
only consider the eigenvalue that is the maximum eigenvalue of 
$ T(x_q,y_q)  {\hat T}(x_q,y_q)$ in the physical regime, so from now on
$ T(x_q,y_q)$, ${\hat T}(x_q,y_q)$, $\tau_j(t_q)$ are to be interpreted as the 
functions for this particular eigenvalue. From (\ref{pfnpsite}) and (\ref{partfn}), noting 
that the the lattice has ${\cal N} = 2 L M$ sites, the
partition function per site is
\be
\kappa_{pq} \eq \lim_{L \rightarrow \infty} \left[ T(x_q,y_q) {\hat T}(x_q,y_q)
\right]^{1/ 2 L} \period \ee

We have distinguished the rapidities $p$, $p'$ of the odd and even vertical rapidity 
lines as it is easy to go from the above equations to those for a model where
the rapidities $p(1), \ldots , p(2L)$ of the $2L$ vertical rapidity lines 
are all different: wherever one sees an expression 
involving $p, p'$ raised to the power $L$, simply replace it by a product over the odd 
and even rapidities, respectively. Thus (\ref{hdef}) becomes
\bd
h_{j,q} \eq \prod_{m=1}^{j-1} \prod_{r=1}^L\left\{ \frac{y_{p_{2r-1}} y_{p_{2r}} 
(x_{p_{2r}} - \omega^m x_q)}
{(y_{p_{2r-1}} - \omega^m x_q )(t_{p_{2r}}- \omega^m t_q ) } \right\} \period \ed

However, from now on we shall only consider the fully homogeneous model, with 
vertical rapidity $p' = p$ for all columns. In this case $T$ and $\hat{T}$ differ 
only in a cyclic shift of the
$L$ spins, and since we are considering only the maximum eigenvalue, the associated 
eigenvector is unchanged by such a shift. This eigenvector is also unchanged by 
multiplication by $X$, {\em so hereinafter we take}
\be p' = p \sep \hat{T}(x_q, y_q) = T(x_q, y_q) \sep X = 1 \period \ee

\section{Functional relations in the infinite-lattice limit}

We emphasize that the above equations are exact for a lattice with a finite number 
$L$ of sites per row. In  \cite{Seoul90, Baxter90, BaxSkew93} they were solved 
in the limit of $L$ large. A key step in this working was to note that in a selected
domain  on the $(x_q, y_q)$ surface, certain terms in each relation became 
exponentially small (relative to the other terms)
as $L$ became large, so could be neglected.

We can identify such terms by examining the low-temperature limit, when $k' \rightarrow 0$ and 
$k \rightarrow 1$.
Similarly to \cite{Seoul90}, take $y_p,  y_q \rightarrow 1$, $x_p \simeq t_p$, $x_q \simeq t_q$, 
$\lambda_p = {\rm O}(k')$, 
$\lambda_q = {\rm O}(k')$. Then $z(t_q) = {\rm O}({k'}^{2L/N})$, and from the definitions 
in \cite{BBP90},
\be \label{kpsmall}  T(x_q, y_q )  \simeq 1 \sep
 \tau_2 (t_q) \simeq (1-\omega t_q)^L  \sep \tau_j(t_q) \simeq {\rm O}(1) \period \ee

It follows that all 
terms in (\ref{tau1}) - (\ref{tau3}) that contain the function $z(t_q)$ or 
$\ab_q$ will be negligible, giving
\be\label{taujeqn}
\tau_j (t_q) \eq \tau_2 (t_q) \tau(\omega t_q) \cdots \tau_2 (\omega^{j-2} t_q) \comma \ee
\be \label{tauprod}
\tau_2 (t_q) \tau_2(\omega t_q) \cdots \tau_2 (\omega^{N-1} t_q) \eq \alpha_q \period \ee

\noindent Similarly, the last term in (\ref{4.20}) will be neglible, so
\be \label{tau2eqn}
\tau_2 (t_q) \eq \left[ \frac{
 (y_p - \omega x_q ) (t_{p} - t_q)}{y_p^2 \, (x_{p} - x_q) } \right]^L 
 \; \frac{ T(x_q, y_q) }{T(\omega x_q, y_q) } \period \ee
Replacing $x_q, y_q, t_q$ in this equation by $\omega^j x_q, y_q, \omega^j t_q$ and 
taking the product over $j = 0, \ldots ,N-1$, we obtain (\ref{tauprod}).

The last term on the rhs of (\ref{3.46}) will also be neglible, 
provided $\lambda_q = {\rm o}({k'}^{(N-2j)/N} ) $, i.e. for $j = 1, \ldots , N$.
(The $j=0$ equation is the same as $j=N$, so we lose nothing by ignoring $j=0$.)

Taking $j=1$, we obtain
\be \label{3.46_1}
T(x_q, y_q) \, T(y_q, \omega x_q)  =  r_{p,q}  = r_{p}(x_q,y_q) \period \ee

Keeping $y_q$ fixed, replacing $j$ by $j-1$ and $x_q$ by $\omega x_q$ in (\ref{3.46}),
then dividing the resulting equation into the original equation (\ref{3.46}),  we get
\be \label{tauasratio}
 \frac{T(x_q,y_q) }{ T(\omega x_q, y_q)} \eq  \left\{ \frac{y_p^2 (x_p - x_q) }
{(y_p - \omega x_q) (t_p - t_q) }  \right\} ^L \tau_2 (t_q) \ee
for $j = 2, \ldots ,N$. But this is the same as (\ref{tau2eqn}), so there is only 
one further equation contained in the set (\ref{3.46}), namely (\ref{3.46_1}).\footnote
{To put this point another way, in the large lattice limit all the $N+1$ identities 
(\ref{3.46}) can be obtained from (\ref{tau2eqn}) and (\ref{3.46_1}).}

To summarize: (\ref{tau2eqn}) can be regarded as defining $\tau_2(t_q)$, 
and (\ref{taujeqn}) as defining $\tau_j(t_q)$. The product relation 
(\ref{tauprod}) is a direct consequence of these definitions. So, out of all the 
functional relations we originally wrote down, the only relation left that contains 
information about the function $T(x_q, y_q)$ is (\ref{3.46_1}).

\subsubsection*{The function $S(\lambda_q)$}

We shall need another function, also introduced in 
\cite{Seoul90},\footnote{This definition differs from that in \cite{Seoul90}, but 
only in an extra  factor $N^{N/2}$ $(-1)^{(N-2)(N-1)/2}$.} namely
\be \label{defS}
S(\lambda_q) \eq \xi_q^L \, T(x_q, y_q) T( \omega x_q, y_q) \cdots
 T(\omega^{N-1} x_q, y_q ) \comma \ee
where
\be \label{defxi}
\xi_q \eq  \prod_{j=1}^{N-1} \left[ \frac{\mu_p \, (
y_{q} - \omega^{-j} y_{p}) (y_q - \omega^{j+1} x_p )}{ \, y_p^2 } \right]^j \period \ee
In \cite{Seoul90, Baxter90} it is shown, for finite $L$, that $S(\lambda_q) $  is 
a polynomial in  $\lambda_q$ of degree $(N-1)L$.

\subsubsection*{Domains}

The parameters $x_q, y_q, t_q, \lambda_q$ are multi-valued functions of one another 
and we have to be careful to identify them when working with the above equations.

\begin{figure}[hbt]  
\begin{picture}(180,240) (-40,-40)
\put(70,100) {\line (1,0) {200}}
\put(170,15) {\line (0,1) {170}}
{\thicklines
\put(115,9){\line(3,5){35}}
\put(115,191){\line(3,-5){35}}
\put(116,9){\line(3,5){35}}
\put(116,191){\line(3,-5){35}}
\put(205,100){\line(1,0){70}}
\put(205,101){\line(1,0){70}}
\thinlines}
\put(237,98){$\bullet$}
\put(131,157){$\bullet$}
\put(131,37){$\bullet$}
\put(158,88){O}
\put(233,88){1}
\put(243,106){$M_0$}
\put(126,150){$\omega$}
\put(132,170){$M_1$}
\put(118,42){$\omega^{-1}$}
\put(136,26){$M_{N-1}$}
\put(0,-25) {Figure 2: The complex $t_q$-plane with its $N$ branch cuts  $M_0,
 \ldots,M_{N-1}$}
\put(120,-42) {(for $N = 3$). }
\end{picture}
\end{figure}

From (\ref{relxy}) and (\ref{deftlambda}), $t_q$ and $\lambda_q$ are related by
\be \label{tlambda}
k^2 t_q^N \eq 1 - k' (\lambda_q + \lambda_q^{-1}) + {k'}^2 \period \ee

Define 
\be \eta = [(1-k')/(1+k')]^{1/N} \sep  0 < \eta < 1 \comma \ee
and consider the cut $t_q$-plane  shown in  \tqplane, with branch points at
$t_q = \omega^j \eta$ and $t_q = \omega^j/ \eta$  and  
a branch cut $M_j$ in between, for $j = 0, \ldots N-1$,as shown.
Then $\lambda_q$ is a single-valued analytic function of $t_q$ in 
this cut  plane, and we can require (consistently with the low-temperature regime
considered above) that $|\lambda_q| < 1$. We can then choose $y_q$ to be the solution of
(\ref{relxy}) such that 
$ |\arg y_q | < \pi/2 N $.
(If $k'$ is small, 
this means that   $y_q \simeq 1$.)  Finally, we define $x_q$ to be
$t_q/y_q$.

The result is that $x_q$ lies in the region $\cal E$ of \xyplane, while
$y_q$ lies in the region ${\cal R}_0$. The boundaries of ${\cal R}_0, \ldots, {\cal R}_{N-1}$
are where  $|\lambda_q| = 1$.

%
\begin{figure}[hbt]
\begin{picture}(420,274) (-40,-74)
\put(70,100) {\line (1,0) {200}}
\put(170,15) {\line (0,1) {170}}
\put(240,100){\circle{60}}
\put(135,161){\circle{60}}
\put(135,39){\circle{60}}
\put(237,97){$\bullet$}
\put(132,158){$\bullet$}
\put(132,36){$\bullet$}
\put(158,88){O}
\put(212,150){${\cal E}$}
\put(230,87){${\cal R}_0$}
\put(126,149){${\cal R}_1$}
\put(121,29){${\cal R}_{N-1}$}
\put(0,-15) {Figure 3: The $N+1$ regions ${\cal E}, {\cal R}_0, \ldots , {\cal R}_{N-1}$ of the complex plane  }
\put(35,-32){in which $x_q$ and $y_q$ lie  (for $N = 3$). ${\cal R}_0, \ldots {\cal R}_{N-1}$ are the}
\put(35,-49){interiors of the approximate circles centred on $1, \omega , \ldots ,\omega^{N-1}$.}
\put(35,-66){ ${\cal E}$ is the the complex plane outside all $N$ circles.}
\end{picture}
\end{figure}

Let us use the symbol ${\cal D}_0$ to denote the domain we have just specified, i.e.
\be {\cal D}_0: \; \; \; \; |\lambda_q|  < 1 \sep x_q \in {\cal E} \sep y_q \in {\cal R}_{0} \period \ee
Since $t_q$ and $\lambda_q$ are uniquely determined if both $x_q$ and $y_q$ are known, 
we shall often
say that ``$(x_q, y_q)$ lies in ${\cal D}_0$'' if the above constraints are satisfied, but it must 
of course be remembered that $x_q$ , $y_q$ are related complex variables.

 If $k'$ is small, $k \simeq 1$ and $(x_q, y_q)$ lies in ${\cal D}_0$, then $y_q \simeq 1$ and $x_q$ is barred only 
from small regions about the $N$ points $1, \omega, \ldots \omega^{N-1}$.

If $x_q$ is an allowed value in ${\cal D}_0$, then so are $\omega^j x_q$, for all integers $j$.
We shall sometimes regard $x_q$ as the independent variable in ${\cal D}_0$.

We shall need to consider the domains (Riemann sheets in the $t_q$-variable) neighbouring 
${\cal D}_0$. There are $N$ of these, obtained
from ${\cal D}_0$ by moving $t_q$ across one of the $N$ branch cuts in  \tqplane. Then
$x_q$ moves into one of ${\cal R}_0, \ldots, {\cal R}_{N-1}$ in \xyplane, while 
$y_q$ moves into $\cal E$.  For instance, if we move
$t_q$ across the cut $M_r$ from $\omega^r \eta$ to $\omega^r /\eta$, we enter the
domain ${\cal D}'_r$, where
\be {\cal D}'_r: \; \; \; \; |\lambda_q| >1  \sep  x_q \in {\cal R}_{r} \sep y_q \in {\cal E} 
 \period \ee
Now it is $x_q$ that is severely restricted,, while $y_q$ is free to move about most 
of the complex plane.

By continuation from the low-temperature, small $k'$, limit, the functional relations
(\ref{taujeqn}) - (\ref{defxi}) hold for $(x_q, y_q)$  and $(x_p, y_p)$ both lying in ${\cal D}_0$, 
except that we must be 
particularly careful with (\ref{3.46_1}), since the arguments $(x_q,y_q)$, $(x_q', y_q') = 
(y_q, \omega x_q)$ of the two  $T$-functions therein cannot both lie in ${\cal D}_0$. One of them 
must lie in a neighbouring domain, but which domain?  The answer is that we want both
$(x_q, y_q)$ and $(x_q', y_q')$ to be in or near the physical regime. Only there can we 
expect to obtain the correct free energy in the large-lattice limit.

If $(x_q, y_q)$ is in  the physical regime, then, from our remarks before (\ref{pfnpsite}),
 $x_q, y_q, t_q$ lie on the unit circle and 
\bd \arg y_q - 2 \pi/N < \arg x_q < \arg y_q  \comma 
\ed
If it also in ${\cal D}_0$, then $y_q$ must be inside ${\cal R}_0$, so $x_q$ must be on 
that portion of the unit circle between between regions ${\cal R}_0$ and ${\cal R}_{N-1}$ in 
\xyplane. Hence $x_q' \in {\cal R}_0$, so $(x_q', y_q')$ lies in ${\cal D}'_0$. Writing
$T_r(x_q,y_q)$ for the analytic continuation of $T(x_q,y_q)$ to domain ${\cal D}'_r$, we 
should therefore write (\ref{3.46_1}) as
\be \label{3.46_1a}
T(x_q, y_q) \, T_0(y_q, \omega x_q) \eq r_{p}(x_q,y_q) \period \ee

Alternatively, $(x_q', y_q')$ may lie in the physical regime  and in ${\cal D}_0$, 
with $y_q' \in {\cal R}_0$. Then $x_q \in {\cal R}_{N-1}$ and, dropping the suffixes on  
$(x_q', y_q')$, (\ref{3.46_1}) becomes
\be \label{3.46_1b}
T_{N-1}(\omega^{-1} y_q, x_q) \, T(x_q, y_q) \eq r_{p}(\omega^{-1} y_q, x_q)  \period \ee

Both (\ref{3.46_1a}) and (\ref{3.46_1b}), with $(x_q, y_q) \in {\cal D}_0$, are valid 
interpretations of (\ref{3.46_1}).

We refer to ${\cal D}_0$  as the {\em principal} or {\em central } 
domain (or Riemann sheet). In a subsequent paper we intend to discuss the full Riemann surface
formed by analytically continuing $\lim_{L\rightarrow \infty} T(x_q,y_q)^{1/L}$  in both the 
$q$ and the $p$ variables.

\section*{The inversion relation}

We now turn from the first to the second general method for calculating free energies, 
the inversion relation method \cite{Stroganov79,Baxter80,Baxter82,book82}.  Let 
\be \label{defDP}
\overline{D}_{pq} = {\rm det}_N \Wb_{pq} (i-j) \sep P_{pq} = \prod_{n=0}^{N-1} W_{pq}(n) \comma
\ee
i.e. $\overline{D}_{pq}$ is the determinant of the $N$ by $N$ Toeplitz (cyclic) matrix with 
entry $\Wb_{pq} (i-j)$ in row $i$ and column $j$. Then under quite general circumstances
the partition function per site $\kappa_{pq}$ satisfies the inversion relation
\be \label{invgen}
\kappa_{pq} \kappa_{qp} \eq \left[ \overline{D}_{pq} P_{pq} \overline{D}_{qp} P_{qp} 
\right]^{1/N} \period \ee

For the chiral Potts model $P_{pq} P_{qp} = 1$ and (\ref{invgen}) becomes, using (2.48) of 
\cite{BBP90},
\be \label{inv1}
\kappa_{pq} \kappa_{qp} \eq r_{pq}^{1/L} \comma \ee
where $r_{pq}$ is defined in (\ref{defrpq}).

Let $R$ be the operator that acts on the rapidity $p$ so that
\be \label{rotn}
x_{Rp} = y_p \sep y_{Rp} = \omega x_p \sep \mu_{Rp} = 1/\mu_p \period \ee
Then replacing  $p,q$ by $q, Rp$ is equivalent to rotating the lattice through
$90^{\circ}$. This does not change the partition function per site, so
\be \label{krotn}
\kappa_{pq} =\kappa_{q,Rp} \period \ee

Combining (\ref{inv1}) and (\ref{krotn}) gives 
\be \kappa_{pq}  \kappa_{p,Rq} = r_{pq}^{1/L} \period \ee

Remembering that $\kappa_{pq} = T(x_q, y_q)^{1/L}$, it follows that
\be \label{inv}
T(x_q,y_q) T(y_q,\omega x_q) = r_{pq} \period \ee

This is {\em precisely} the relation (\ref{3.46_1}), again with one function $T$ 
being understood to be the analytic
continuation of the other from the physical regime. Thus we did not need to go through
all the working that was necessary in \cite{BazStrog90}, \cite{BBP90} to derive 
the functional relations (\ref{3.46}) - (\ref{tau01}). We could have used the simple
inversion relation, obtained (\ref{inv}), i.e. (\ref{3.46_1}), then defined the 
auxiliary functions $\tau_j(t_q)$, $S(\lambda_q)$ so as to obtain
the infinite-lattice  relations (\ref{taujeqn}) - (\ref{defxi}) above.

\section{Solution of the infinite-lattice functional relations}

We now solve the infinite-lattice relations (\ref{taujeqn}) - (\ref{defxi}) successively for
$\tau_2(t_q)$, $S(\lambda_q)$, $T(x_q,y_q)$, thereby obtaining the partition function per site
$\kappa_{pq}$. Each side of every one of these relations is a quantity raised to the power $L$, 
and this is the only way $L$ enters the relations. We should define new functions that are the $L$th
roots of $\tau_2(t_q)$, $S(\lambda_q)$, $T(x_q,y_q)$, and write down the $L$th roots of each 
relation. Equivalently, from now on we formally set 
\be L = 1   \ee
in equations (\ref{defz}) - (\ref{hdef}) and (\ref{kpsmall}) - (\ref{inv}).

Thus
$T(x_q, y_q)$, or $T_{pq}$,  is now the partition function per site $\kappa_{pq}$ of the lattice, 
defined as in the paragraph containing (\ref{pfnpsite}), (\ref{freeenergy}), i.e.
\be \kappa_{pq}  = T_{pq} = T(x_q,y_q) \period \ee

Our derivation will be based solely on the inversion relation (\ref{3.46_1}), as interpreted in 
(\ref{3.46_1a}), and the  definitions in
(\ref{taujeqn}) - (\ref{defxi}). It will necessarily involve various analyticity
assumptions, and in making these we have been heavily guided by the previous derivations in
\cite{Seoul90}, \cite{Baxter90}. In particular, we shall regard $\tau_2(t_q)$ as a function of
$t_q$ (rather than, say, $x_q$ or $\mu_q$), and  $S(\lambda_q)$ as a function of $\lambda_q$. It 
must be admitted that from the present point of view this is by no means an obvious thing to 
do. We shall present what justifications we can, and attempt to state clearly the analyticity
assumptions that we make.

In particular, the domains ${\cal D}_0, {\cal D}'_0, \ldots, {\cal D}'_{N \! - \! 1}$ are 
made up of  $N+1$ connected Riemann sheets of the cut $t_q$ plane of  \tqplane. 
Together they are
the beginnings of the full Riemann surface on which the functions live (and which we intend to 
discuss in a later paper). Here we only need these domains or sheets. In fact we only need the surface 
consisting of ${\cal D}_0$ and the adjacent neighbourhoods of ${\cal D}'_0, \ldots, 
{\cal D}'_{N \! - \! 1}$
(obtained by just crossing the branch cuts in  \tqplane). Since $|\lambda_p| < 1$, and on 
${\cal D}_0$
it is true that $|\lambda_q| < 1$, we can choose this surface so that
\be \label{defDPlus}
|\lambda_p \lambda_q | < 1  \; \; \; \; {\rm on} \; \; \; {\cal D}_{+} \period \ee
  Let us call this extended surface ${\cal D}_{+}$. It includes the central domain ${\cal D}_0$ and all functions are defined on it by analytic continuation from
 ${\cal D}_0$.  
Then a basic assumption that we make is:

\vspace{0.8cm}

ASSUMPTION 1: {\em The function $T(x_q,y_q)$ is non-zero and analytic
on ${\cal D}_{+}$.}

\vspace{0.8cm}

This is consistent with the low-temperature result and is a standard assumption in the inversion
relation method: the free energy is analytic in some fundamental domain that includes 
the physical regime. The remark after equation (\ref{partfn}) that $\mu_q$ enters $T(x_q,y_q)$
only via its $N$th power $\lambda_q$ is important here: it it were not true then there would
also be branch 
cuts at $0$ and $\infty$ in the complex $t_q$ plane, and the surface would 
be even more complicated.

In \cite{Seoul90, Baxter90} we took $L$ to be large but finite and used the fact that
$\tau_j(t_q), S(\lambda_q)$  are then polynomials to solve the functional relations. Here
we adopt a different (but related) strategy: we use Weiner-Hopf factorization to solve 
(\ref{taujeqn}) - (\ref{defxi}).

\subsubsection*{Calculation of $\tau_2 (t_q)$}

Since ${\cal D}_{+}$ extends beyond the branch cuts in Fig. 2, $\tau_2 (t_q)$
is not necessarily a single-valued function of $t_q$: we may expect it to have these branch 
cuts in the complex $t_q$ -plane.

First consider possible poles or zeros.
From Assumption 1 and (\ref{tau2eqn}), these can occur only when 
  $\lambda_q = \lambda_p$, $x_q^N = x_p^N$ and  $y_q^N = y_p^N$, or when 
  $\lambda_q = 1/\lambda_p$, $x_q^N = y_p^N$ and  $y_q^N = x_p^N$. The second possibility is 
excluded by (\ref{defDPlus}), so we have to consider the first. 
Since $y_p, y_q \in {\cal R}_0$, this can only happen when
$y_q = y_q$. But then the term $(t_p-t_q)/(x_p-x_q)$ simplifies to $y_p$: the potential 
pole and zero cancel one another. Thus 
 $\tau_2(t_q)$ is analytic and non-zero in  ${\cal D}_{+}$. From (\ref{taujeqn}), so 
therefore is $\tau_N(t_q)$.

From (\ref{taujeqn}) and (\ref{tau2eqn}),
\be
\tau_N(t_q) \eq \frac{(x_p-\omega^{-1} x_q) (y_p^N-x_q^N) (t_p^N-t_q^N) \; T(x_q,y_q)}
{y_p^{2N-2} \, (x_p^N- x_q^N) (y_p-x_q) (t_p - \omega^{-1} t_q) \; 
   T( \omega^{-1} x_q, y_q)} \period \ee 

\noindent Also, from (\ref{defrpq}) and (\ref{3.46_1a}),
\be \frac{1}{T(\omega^{-1} x_q, y_q)}  \eq \frac{(x_p^N-x_q^N) (y_p^N - y_q^N) 
(t_p - \omega^{-1} t_q) \; T_0(y_q, x_q) } {N (x_p - \omega^{-1} x_q) 
(y_p - y_q) (t_p^N-t_q^N) } \period   \ee

\noindent Combining these together, there are several cancellations, leaving
\be \label{Tprod}
\tau_N(t_q) \eq \frac{ (y_p^N-x_q^N) (y_p^N-y_q^N)  \; T(x_q,y_q) \, T_0(y_q,x_q)}
{N \, y_p^{2N-2} \, (y_p-x_q) (y_p-y_q)   \; } \period \ee 

Now $T_0(x_q,y_q)$ is the analytic continuation of $T(x_q,y_q)$ across the cut $M_0$, where 
$x_q$ and $y_q$ are both on the boundary of ${\cal R}_0$.
Analytically continuing the rhs of (\ref{Tprod}) across this cut and then interchanging
$x_q$ with $y_q$ leaves it unchanged, while also leaving $t_q$ unchanged. For a given
$t_q$, these are the only possible values of $(x_q, y_q)$ in this neighbourhood.
Thus in the neighbourhood of $M_0$, on either side of the cut, 
$\tau_N(t_q)$ is a single-valued function of $t_q$. There is therefore no need for the 
cut $M_0$ for this function: it can be removed.

From (\ref{taujeqn}),  the equation (\ref{tauprod}) can be written as
\be \label{tauN2eqn}
\tau_N(t_q) \tau_2(\omega^{-1} t_q) \eq \alpha_q \period \ee
The rhs of this equation is certainly is not a single-valued function of $t_q$ across the cut
$M_0$ (when $\lambda_q \rightarrow 1/\lambda_q$), so nor is
$\tau_2(\omega^{-1} t_q)$. Hence the plane in which $\tau_2 (t_q)$ is analytic
must have $M_{N\minus 1}$ as a branch cut.

\vspace{0.8cm}

ASSUMPTION 2: {\em $\tau_2 (t_q)$  is single-valued, with logarithmic derivative  
zero at infinity,  in the $t_q$ plane containing only the cut $M_{N \minus 1}$. }

\vspace{0.8cm}

This is by no means obvious. It is consistent with $\tau_N(t_q) = $ 
$\tau_2(t_q) \cdots $ $\tau_2(\omega^{N-2} t_q)$
not having the cut $M_0$, but is not implied by it. It is possible
that  $\tau_2 (t_q)$ could have cuts other than  $M_{N \minus 1}$ that cancelled one another
in the product function  $\tau_N(t_q)$. On the other hand, it is the simplest assumption 
consistent with  $\tau_N(t_q)$ 
not having $M_0$ as a branch cut, and again part of the spirit of the inversion relation method
is to assume the simplest possible analytic structure.\footnote{We of course know that 
Assumption 2 is true
from \cite{Seoul90}, but are trying to present an argument based  on the inversion 
relation.} Exhibiting
the possible  multi-valuedness of $\tau_2(t_q)$ by writing it as $\tau_2(x_q,y_q)$, our 
assumption  implies that for  $(x_q,y_q)$ on ${\cal D}_{+}$,
\be \label{tau2symm}
\tau_2(x_q,y_q) \eq \tau_2(\omega^{r} y_q,\omega^{-r} x_q) \ee
for $r = 0, \ldots, N-2$. So we are assuming these additional symmetries.

The requirement that the logarithmic derivative be zero at infinity follows from 
the finiteness and 
analyticity of the Boltzmann weights and hence $T(x_q,y_q)$ when $x_q \rightarrow \infty$ 
in $\cal E$.

It follows from these two assumptions and the above remarks 
that there is a closed curve $\cal C$ is the complex $t_q$ plane, surrounding the
potential branch points $\eta$, $1/\eta$ as in  \tqplane, such that  

\noindent i)   $ \tau_N( t_q)$  is analytic and non-zero inside and on $\cal C$,

\noindent ii)  $ \tau_2( \omega^{-1} t_q)$ is analytic and non-zero  outside and on $\cal C$,

\noindent iii)  $ (d/dt_q) \log \tau_2( \omega^{-1} t_q) \rightarrow 0$ as $t_q \rightarrow \infty$.

We can now solve (\ref{tauN2eqn}) by Wiener-Hopf factorization.\cite{Noble58} Regard 
$\alpha_q$ as a function $\alpha (t_q)$ of $t_q$ and temporarily drop the suffix $q$. Define
\ba \label{defF}
F_{-}(s) & =  & - \frac{1}{2 \pi i} \oint_{{\cal C}} \frac{1}{t-s} \left(
\frac{d}{d t} \log \alpha (t) \right) dt \sep {\rm  s \; \; outside \; \;  } {\cal C}
\comma  \\
F_{+}(s) & = &  ~~ \frac{1}{2 \pi i} \oint_{{\cal C}} \frac{1}{t-s} \left( \frac{d}{d t} 
\log \alpha (t) \right) dt  \; \sep {\rm  s \; \; inside \; \;  } {\cal C} \period \nonumber \ea

Shifting the  contour for $F_{-}(s)$ to be inside $\cal C$, that for $F_{+}(s)$  to be outside, 
and $s$ inbetween, it follows from Cauchy's integral formula that
\be
F_{-}(s) + F_{+}(s) \eq  \alpha'(s)/\alpha(s) \period \ee

Hence from (\ref{tauN2eqn}),
\be
F_{-}(s)  - \frac{d}{ds} \log \tau_2(\omega^{-1} s) \eq - F_{+}(s)  +
 \frac{d}{ds} \log \tau_N( s)  \period \ee
The rhs of this equation is analytic for $s$ inside and on $\cal C$, while the lhs
is analytic outside and on $\cal C$ and tends to zero as $s \rightarrow \infty$. Hence 
both sides are entire and bounded. By Liouville's theorem they must each be constant, 
in fact zero, so
\be \label{dtaus}
\frac{d}{ds} \log \tau_2(\omega^{-1} s) = F_{-}(s) \period \ee

We can shrink $\cal C$ to just surround the branch cut in the $t$ plane from $\eta$ to
$1/\eta$. Then as $t$ goes round $\cal C$, $\lambda = \lambda_q$ (as defined by (\ref{tlambda}), goes 
anti-clockwise round the unit circle. Changing the variable of integration in (\ref{defF}) to 
$\theta$, where $\lambda = e^{i \theta}$, then replacing $s$ by $\omega t_q$, it follows that
\be \label{tau2result}
\log \tau_2 (t_q) \eq \frac{1}{2\pi} \, \int_{0}^{2 \pi} \left( \frac{1 + \lambda_p e^{i \theta}}
{1 - \lambda_p e^{i \theta}} \right) \, \log \! \left[ \frac{\Delta(\theta) - \omega t_q }{y_p^2} 
\right] \, d\theta \ee
for $t_q$ lying in the complex plane with a single branch cut from $\omega^{-1} \eta$ to 
$\omega^{-1} /\eta$. Here
\be  \label{defDel}
\Delta (\theta) \eq \left( \frac{1 -2 k' \cos \theta + {k'}^2}{k^2} \right)^{1/N} \ee
and we have integrated (\ref{dtaus}). We have fixed the integration constant
by  using (\ref{tauprod}) and the formula
\be \frac{1}{2\pi} \, \int_{0}^{2 \pi} \left( \frac{1 + \lambda_p e^{i \theta}}
{1 - \lambda_p e^{i \theta}} \right) \, \log [\Delta(\theta)^N -  t_q^N ]  \eq \log 
\left[ \frac{k' (1-\lambda_p \lambda_q)^2}{k^2 \lambda_q} \right ] \comma \ee
which is true when $t_q, \lambda_q$ are related by (\ref{tlambda}) and 
$|\lambda_p| < 1 $, $|\lambda_q| < 1 $. It also follows from this formula
and (\ref{tauN2eqn}) that $\tau_N(t_q) = \tau_2(t_q) \cdots \tau_2(\omega^{N-2} t_q)$, 
in agreement with (\ref{taujeqn}).

\subsubsection*{Calculation of  $S(\lambda_q)$}

The function $S(\lambda_q)$ is defined by (\ref{defS}) {\em in the domain} ${\cal D}_0$, where 
$|\lambda_q| $ is sufficiently small as to justify our neglect of various
terms in the functional relations in the limit of $L$ large. (At low temperatures, 
this means that $|\lambda_q|  < k'$). Outside this 
domain we here define it by analytic continuation. 

This is a 
different definition from \cite{Seoul90}, where much use was made of the fact that $S(\lambda_q)$
is a polynomial for finite $L$, with zeros located approximately on  concentric circles between
$|\lambda_q| = k'$ and  $|\lambda_q| = 1/k'$. In the large-$L$ limit these cause 
the function $S(\lambda_q)$ of \cite{Seoul90} to have a different analytic form across 
each such circle of zeros, so the two definitions will only agree for $|\lambda_q| $ inside 
the smallest circle.

We first repeatedly use (\ref{tau2eqn}) to express $T(\omega x_q, y_q), \ldots ,
T(\omega^{N-1} x_q, y_q)$ in (\ref{defS}) in terms of $T( x_q, y_q)$, giving
\be \label{defS2}
S(\lambda_q) \eq \xi_q \, T(x_q,y_q)^N \prod_{j=1}^{N-1} 
\left\{ \frac{(y_p-\omega^j x_q)(t_p - \omega^{j-1} t_q) }
{y_p^2 \, (x_p - \omega^{j-1} x_q) \, \tau_2 (\omega^{j-1} t_q) } \right\}^{N-j} \period \ee

Within ${\cal D}_0$, if $\lambda_q = \mu_q^N$ is given, we take $y_q$ to be the root of (\ref{relxy})
lying in ${\cal R}_0$. There are $N$ possible choices for $x_q$ in $\cal E$, but by
the construction  of (\ref{defS}), $S(\lambda_q)$ is the same for each, so it is a single-valued
function of $\lambda_q$ in ${\cal D}_0$ and (by analytic continuation) in ${\cal D}_{+}$.  
From Assumption 1 and the definition (\ref{defS}),
{\em it is analytic and non-zero in ${\cal D}_{+}$}. 

For $|\lambda_q|$ close to one, choose  $x_q$, like $y_q$, to lie near the boundary of 
${\cal R}_0$.
Replacing $\lambda_q$ by $1/\lambda_q$ simply interchanges $x_q$ with $y_q$, while leaving $t_q$ unchanged. Note that
$\tau_2(\omega^{-1} t_q)$ does {\em not} occur on the rhs of (\ref{defS2}), so allowing
$\lambda_q$ to move just outside the unit circle does not take any of the $\tau_2$ functions
in (\ref{defS2}) across a branch cut in the $t_q$ plane and they are
unchanged by  analytically  continuing  from $\lambda_q$ to $1/\lambda_q$.

The function $T(x_q, y_q)$ is simply replaced by 
$T_0(y_q, x_q)$, which is given by (\ref{Tprod}). So analytically  continuing  (\ref{defS2})
from $\lambda_q$ to $1/\lambda_q$ (thereby interchanging $x_q$ with $y_q$), then multiplying 
by the original equation, we obtain (after many cancellations)
\be \label{Sl1l}
S(\lambda_q) S(1/\lambda_q) \eq N^N \, \left( \lambda_p /y_p^{2N} \right)^{N-1} \; \prod_{j=1}^{N-1} 
 \frac{
(t_p - \omega^{j-1} t_q)^{2N-2j} }
{ \tau_2 (\omega^{j-1} t_q)^{N-2j} }  \period \ee
This is {\em not} the same as equation (26) of \cite{Seoul90}, because here $ S(1/\lambda_q) $ is 
defined by analytic continuation of $ S(1/\lambda_q) $ from $|\lambda |<1 $ to $|\lambda| > 1 $. As 
remarked above, this is a different definition from that used in \cite{Seoul90}.

The rhs is  known. As $\lambda_q$ moves round the unit circle, $t_q$ moves 
on the positive real axis from $\eta$ to $1/\eta$ and back again. The logarithm of the rhs
is analytic on this contour, returning to its original value. It follows from the above 
remarks that
{\em  $\log S(\lambda_q)$ is analytic inside and on the unit circle in the complex
$\lambda_q$-plane.}

We can therefore  solve
(\ref{Sl1l}) for $S(\lambda_q)$ by Wiener-Hopf factorization.
Writing the rhs of (\ref{Sl1l}) as $R[t_q]$,  we obtain
\be \log S(\lambda_q) \eq \frac{1}{4\pi} \int_0^{2 \pi} \frac{1+\lambda_q e^{i \theta} }
{1-\lambda_q e^{i \theta} } \, \log R[\Delta (\theta)] \comma \ee
where $\Delta(\theta)$ is defined by (\ref{defDel}). Define,\footnote{These definitions are the same as those of eqns. (45) and (46) 
of \cite{Seoul90}, except for the factor $k^{2/N}/(1- \omega^j)$ inside the logarithm in  
$A_{pq}$. This gives an extra
constant contribution $(N-1) \log k + $ $ i \pi (N-2)(N-1)/12 - (N/2) \log N$ to $A_{pq}$, 
and ensures that $A_{pq} = 0$ in the low-temperature limit of the Appendix.} for $|\lambda_p| < 1$ and 
$|\lambda_q| < 1$: 
\ba \label{defAB}
A_{pq} & = & \frac{1}{2 \pi} \int_0^{2 \pi} \frac{1+\lambda_p e^{i \theta} }
{1-\lambda_p e^{i \theta} } \, \sum_{j = 1}^{N-1}  (N-j) \log \left\{ \frac{k^{2/N} [ \Delta(\theta) -
\omega^j t_q]}{1-\omega^j} \right\} \, d \theta \comma \\
 B_{pq} & = &\frac{1}{8 \pi^2} \int_0^{2 \pi} \int_0^{2 \pi}
\frac{1+\lambda_p e^{i \theta} }
{1-\lambda_p e^{i \theta} } \,
\frac{1+\lambda_q e^{i \phi} }
{1-\lambda_q e^{i \phi} } \, \nonumber \ea
\be \times \sum_{j=1}^{N-1} (N-2j) \log [\omega^{-j/2} \Delta(\theta) -
\omega^{j/2} \Delta(\phi)  ] \, d\theta \, d\phi \comma \ee
it follows that
\be\label{Sresult}
\log S(\lambda_q) \eq  (N-1) \log \left[ \frac{k' 
(1-\lambda_p \lambda_q)^2} {k \, y_p^N \lambda_p^{1/2} }\right]  
 - A_{qp} - B_{pq} \period \ee

Let $\overline{q} =  \{ y_q, x_q, 1/\lambda_q, t_q , 1/\mu_q\}$, so 
$\lambda_{\overline{q}} = 1/\lambda_q$ and $t_{\overline{q}} = t_q$.
 One can verify that (\ref{Sresult})  does indeed satisfy (\ref{Sl1l}) by 
analytically continuing
$A_{qp}, B_{pq}$  to $|\lambda_q| > 1$  (deforming the contours of integration accordingly),
and verifying that the resulting functions satisfy
\be
A_{qp} + A_{\overline{q} p} \eq 2 \sum_{j=1}^{N-1} (N-j) \, 
\log \left[ \frac{k^{2/N} (t_q - \omega^j t_p) }{ 1- \omega^j } \right] \comma \ee
\be
B_{pq} + B_{p, \overline{q} } \eq  \sum_{j=1}^{N-1} (N-2 j) \, 
\log \left[ \frac{\tau_2(\omega^{j-1} t_q) }{ 1- \omega^j } \right] \comma \ee
provided that for $\lambda_q$ near (on) the unit circle, $t_q$ is taken to be the solution of 
(\ref{tlambda}) near (on) the real positive axis, so that $|\arg (t_q)| < \pi/N$.

\subsubsection*{Calculation of $T(x_q,y_q)$}

Now we know $S(\lambda_q)$, we can solve (\ref{defS2}) for $\kappa_{pq} = T(x_q,y_q)$.
The result is best expressed in terms of $T(x_q,y_q)/(\overline{D}_{pq} P_{pq})^{1/N}$, where
$\overline{D}_{pq}$, $ P_{pq}$ are defined in (\ref{defDP}). From (2.44) of \cite{BBP90},
\be \label{Dbar} 
\overline{D}_{pq} \eq \prod_{j=1}^{N-1} \left[ \frac{(1-\omega^j)(t_p - \omega^j t_q)}
{(x_p - \omega^j x_q)(y_p - \omega^j y_q) } \right]^{j} \comma \ee
while from (\ref{wts}),
\be \label{Pprod}
 P_{pq} \eq (\lambda_p/\lambda_q)^{(N-1)/2} \, \prod_{j=1}^{N-1} \left[ \frac{
y_q - \omega^j x_p }{ y_p - \omega^j x_q } \right]^{N-j} \period \ee

Using these formulae and (\ref{defxi}), (\ref{defS2})  gives
\ba \label{TpowerN}
\frac{T(x_q,y_q)^N}{\overline{D}_{pq} P_{pq}} & = &   
(\lambda_q^{1/2} \alpha_q )^{1-N} \, S(\lambda_q) 
 \prod_{j=1}^{N-1}
\left[ \frac{\tau_2 (\omega^{j-1} t_q) }{1-\omega^j } \right] ^{N-j} \comma  \nonumber \\
 & = &  S(\lambda_q) \, \exp (A_{pq}) \, \left[ \frac{k \, y_p^N \lambda_q^{1/2} }
{k' (1-\lambda_p \lambda_q)^2 } \right]^{N-1} \comma \nonumber  \\
& = & (\lambda_q/\lambda_p)^{(N-1)/2} \, \exp (A_{pq} - A_{qp} - B_{pq} ) \period \ea
This is the result (50) of \cite{Seoul90}.

There are two ``inversion'' relations: the inversion relation (\ref{inv1}) and 
the rotation symmetry (\ref{krotn}). We have derived  (\ref{TpowerN}) from the single
combined relation (\ref{inv}). However, 
 $B_{pq} = -B_{qp}$, so the rhs of  (\ref{TpowerN})  is inverted by interchanging $p$ with $q$.
Thus  (\ref{inv1}) is indeed satisfied
and we have in fact satisfied both relations.

\section{Analytic continuation of $T(x_q,y_q)$}

We can readily use the above equations to obtain the analytic continuation of $T(x_q,y_q)$
from ${\cal D}_0$ to the neighbouring domains ${\cal D}'_0, \ldots ,{\cal D}'_{N-1}$. As in 
section 3,
if $(x_q,y_q) \in {\cal D}_0$ we write $T(x_q,y_q)$ simply as $T(x_q,y_q)$, while if
$(x_q,y_q) \in {\cal D}'_r$ (for $r = 0, \ldots , N \! - \! 1$) we write $T(x_q,y_q)$  
as $T_r(x_q,y_q)$.

First note by repeated use of (\ref{tau2eqn}) that, for $(x_q,y_q)$ in ${\cal D}_0$ and 
$r = 0,\ldots, N \! - \! 1$, 
\be \label{Ttau1}
T(\omega^r x_q, y_q) \eq T(x_q,y_q) \, \prod_{j=1}^{r} \frac{(y_p-\omega^j x_q)
(t_p - \omega^{j-1} t_q) }{y_p^2 (x_p - \omega^{j-1} x_q) \tau_2 (\omega^{j-1} t_q) } \period \ee

Now analytically continue $(x_q,y_q)$ from ${\cal D}_0$ to ${\cal D}'_0$, so that $t_q$ goes 
through the branch cut $M_0$. Then $(\omega^r x_q, y_q) $ 
on the rhs goes from  ${\cal D}_0$ through the cut $M_r$ to ${\cal D}'_r$. The function
$T(x_q,y_q) $ on the rhs therefore becomes the function $T_0(x_q,y_q) $, while 
$T(\omega^r x_q, y_q)$ on the rhs becomes $T_r(\omega^r x_q, y_q)$.

Going through the branch cut $M_0$ and returning to the original value of $t_q$ is equivalent
to interchanging $x_q$ with $y_q$. (The branch points of $M_0$ are where $x_q = y_q$.) We can therefore
interchange $x_q$ with $y_q$ and obtain the equation
\be \label{Ttau2} T_r(\omega^r y_q, x_q) \eq T_0(y_q,x_q) \, \prod_{j=1}^{r} \frac{(y_p-\omega^j y_q)
(t_p - \omega^{j-1} t_q) }{y_p^2 (x_p - \omega^{j-1} y_q) \tau_2 (\omega^{j-1} t_q) } \period  \ee
Like (\ref{Ttau1}), this equation  holds for $(x_q,y_q)$ in ${\cal D}_0$ and  $r = 0, \ldots , 
N \! - \! 1$.
The functions  $\tau_2(t_q) , \ldots ,\tau_2(\omega^{r-1}t_q) $ in these equations are 
single-valued and
analytic functions of $t_q$ across the cut $M_0$, i.e. $M_0$ is not a brach cut of these 
functions. Hence they are the 
same functions, with the same values, in each equation.

We can therefore eliminate the $\tau_2$ functions by dividing (\ref{Ttau2}) by 
(\ref{Ttau1}). Using also the form (\ref{3.46_1a}) of the inversion relation , 
we obtain, for $(x_q,y_q)$ in ${\cal D}_0$ and $r = 0, \ldots , N-1$,
\ba \label{Trcont}
T_r(\omega^r y_q, x_q) & =  & \frac{N \, T(\omega^r x_q,y_q)}{T(\omega^{-1} x_q,y_q) T( x_q,y_q)} \prod_{j=1}^r \frac{t_p - \omega^{j-1} t_q}{
(x_p \!  -  \!  \omega^{j-1} y_q) (y_p  \! -  \! \omega^j x_q) } \; \nonumber \\ 
& & \times \prod_{j=r+1}^{N-1} \frac{t_p - \omega^{j-1} t_q}{
(x_p - \omega^{j-1} x_q) (y_p - \omega^j y_q) }
 \period \ea

Given $T(x_q,y_q)$ in ${\cal D}_0$, this relation enables us to calulate the function
in each neighbouring domain ${\cal D}'_r$. If we set $r = 0$ we regain the inversion relation
(\ref{3.46_1a}), while if
we set $r = N-1$ we obtain the alternative form (\ref{3.46_1b}) predicted in section 3.

One can continue. If we allow $x_q$ in (\ref{Trcont}) to move from ${\cal E}$ into ${\cal R}_s$
(with $s  = 1, \ldots , N-1$), 
then $(x_q', y_q') = (\omega^r y_q, x_q) $ on the rhs  moves into a new domain ${\cal D}_{rs}$, 
where $x_q' \in {\cal E}$ and $y_q' \in {\cal R}_s$. Correspondingly, $T_r(x_q',y_q')$ 
on the lhs becomes the function $T_{rs}(x_q',y_q')$. On the rhs, the $T$ functions become 
$T_{r+s}, T_{s-1}, T_s$, all of which can in turn be expressed in terms of the original function
$T$ (in domain ${\cal D}_0$) by using (\ref{Trcont}).

The we can continue $x_q'$ into ${\cal R}_t$ to obtain the function  $T_{rst}$ whose 
arguments lie in the domain ${\cal D}'_{rst}$, and so on. If each function were unique, then 
this procedure would give a Cayley tree of functions and domains, each having $N$ neighbours,
with no circuits. This would form an infinite-dimensional Bethe lattice. In a subsequent paper 
we intend to show that in fact the functions are not unique and
there are circuits, so that one obtains instead an $N$-dimensional lattice (for $N > 2$), each site 
corresponding to a domain and a function. This is the Riemann surface on which $T(x_q, y_q)$ lives.

By contrast, the Boltzmann weights and related quantities of course form a zero-dimensional 
surface: for instance,
the function ${\overline D}_{pq}$ in (\ref{Dbar}) has just $2N$ sheets in the $t_q$ plane, corresponding
to either $x_q$ or $y_q$ being in ${\cal R}_r$ (for $r = 0, \ldots , N-1$), the other in $\cal E$.

We also hope to explore the question of whether one can describe the Riemann surface for 
$T(x_q,y_q)$  by 
an appropriate generalization of the hyperelliptic
variables and functions defined in \cite{Baxter91}.

\section{Analytic continuation in $x_p,y_p$}

Up till now we have held the $p$ variables (corresponding to the vertical rapidity)
fixed at some values in ${\cal D}_0$,  except possibly in the inversion relations 
(\ref{invgen}) - (\ref{rotn}). One can of 
course vary the $p$ variables  as well as the $q$ variables, and this would be 
necessary for a full understanding of the Riemann surface
on which then free energy lives. In particular, we should like to obtain the $p$-variable 
analogue of (\ref{Trcont}).

From this point of view $T$ is a function $T(x_p,y_p| x_q,y_q)$ of both the $p$ and $q$ variables.
It has an unexpected ratio property. From  (\ref{tau2result}) it is apparent that $\tau_2(t_q)$ 
is unchanged by replacing $x_p, y_p$ by $\omega x_p, y_p$ (the variables remain 
in ${\cal D}_0$ and $\lambda_p$ is unaltered). From
(\ref{tau2eqn}) it follows that, for $(x_p,y_p)$ and $(x_q,y_q)$ both in ${\cal D}_0$,
\be \label{2ndratio}
\frac{T(x_p, y_p| x_q, y_q) \, T(\omega x_p, y_p| \omega x_q, y_q)}{
T(x_p, y_p|\omega  x_q, y_q) \, T(\omega x_p, y_p|  x_q, y_q)} \eq 
\frac{(x_p - x_q)(\omega t_p - t_q)}{(\omega x_p - x_q)( t_p - t_q)} \period \ee
This result is not at all obvious from first principles: it could serve as a useful test of
series expansions.

Note that one has to be very careful in applying the equations of this paper in 
domains other than those
for which they were obtained. This relation is a good example: if we analytically continue
$T(x_p,y_p| x_q,y_q)$ to $(x_p,y_p) \in {\cal  D}_0$,  $(x_q,y_q) \in {\cal D}'_r$ by allowing
$x_q$ to move from $\cal E$ to ${\cal R}_r$, $y_q$  from ${\cal R}_0$ to ${\cal E}$, then
the $T$-functions with third argument $x_q$ will become the functions $T_r$ above, while those
with third argument $\omega x_q$ will become $T_{r+1}$. The different functions, originally 
on the same sheet, will move onto different sheets.

For the rest of this section we  shall hold $x_q, y_q$ fixed, lying in ${\cal D}_0$, so 
for brevity we write 
$T(x_p,y_p; x_q,y_q)$ as $T[x_p,y_p]$.

Fortunately it is not necessary to repeat all the above working.  We can  replace $x_p, y_p$ 
in the final  result 
(\ref{TpowerN})
by $\omega x_p, y_p$ (so again the variables $x_p,y_p$ remain in ${\cal D}_0$) and take the ratio of 
the equation with these arguments to that with the 
original arguments. Since $\lambda_p$ is the same in both, the factors involving $A_{pq}, B_{pq}$ 
cancel, and we obtain
\be \label{tau2pdef}
\hat{\tau}_2(t_p)  \eq \frac{(t_q-\omega t_p)(y_q - \omega x_p)}
{y_q^2 (x_q - \omega x_p) } \; \frac{T[\omega x_p, y_p]}{T[x_p,y_p]} \comma \ee
where $\hat{\tau}_2(t_p)$ is the function $\tau_2(t_q)$ defined by (\ref{tau2result}), but with
$p$ interchanged with $q$, i.e. 
\be \label{tau2presult}
\log {\hat \tau}_2 (t_p) \eq \frac{1}{2\pi} \, \int_{0}^{2 \pi} \left( \frac{1 + \lambda_q e^{i \theta}}
{1 - \lambda_q e^{i \theta}} \right) \, \log \! \left[ \frac{\Delta(\theta) - \omega t_p }{y_q^2} 
\right] \, d\theta \period \ee
It is analytic in the $t_p$ plane, except for a single branch cut from from $\omega^{-1} \eta$ to
$\omega^{-1} \eta^{-1}$. This result is of course consistent with (\ref{2ndratio}).

Let $T_r[x_p,y_p]$ be the analytic continuation of $T[x_p,y_p]$  from
$(x_p,y_p) \in {\cal D}_0$  to  $(x_p,y_p) \in {\cal D}'_r$  (i.e. $x_p$ moves to the region 
${\cal R}_r$ of \xyplane,  $y_p$ to the region ${\cal E}$).
Then from the inversion relations (\ref{invgen}) - (\ref{rotn}),
\be \label{invp0}
T[\omega^{-1}  y_p,x_p] \, T_0[x_p, y_p] \eq r_{pq} \eq
T_{N-1} [\omega^{-1} y_p, x_p] \, T[x_p,y_p] \period \ee
(The second of these relations is the analytic contination of the first.)

Proceeding as in the previous section, from (\ref{tau2pdef}) and either equation of (\ref{invp0}),
 we can deduce that 
\ba \label{Trcontp}
T_r[\omega^r y_p, x_p] & =  & \frac{N \, T[\omega^r x_p,y_p]}{T[\omega^{-1} x_p,y_p] \, T[ x_p,y_p]} 
\; \prod_{j=1}^r \frac{t_q - \omega^{j} t_p}{
(x_q \!  -  \!  \omega^{j} x_p) (y_q  \! -  \! \omega^j y_p) } \; \nonumber \\ 
& & \times \prod_{j=r+1}^{N-1} \frac{t_q - \omega^{j} t_p}{
(x_q - \omega^{j} y_p) (y_q - \omega^j x_p) }
  \ea
for $ r = 0, \ldots , N-1$. 

If $r=0$ we regain the first inversion relation (\ref{invp0}); if $r = N-1$ we 
obtain the second.

\section{Summary}

A basic difficulty with the $N$-state  chiral Potts model is that it lacks 
the ``rapidity-difference'' property \cite{Baxter90}. Because of this,  there is no known explicit parametrization
in terms of elliptic or other single-valued single-variable functions. The author did
show that one could parametrize the Boltzmann weights in terms of hyperelliptic functions with 
$N-1$ related arguments \cite{Baxter91}, but so far little progress has been made using 
this parametrization. (For $N=3$ it has been used to obtain the $f_{pq}$. \cite{Baxter98b}) 
For this reason, it is difficult to see how to use the 
standard, usually very simple, ''inversion relation'' method \cite{Stroganov79, Baxter82}
to obtain the free energy.

Here we have addressed this problem.  We have shown that the inversion and 
rotation relations (\ref{inv1}) and 
(\ref{krotn}) give  (\ref{inv}), and that this is sufficient to deduce the ``infinite 
lattice functional relations'' (\ref{taujeqn}) - (\ref{defxi}), which from this point of view are
simply definitions and direct consequences of the definitions.

We then made two assumptions as to the analyticity properties of the functions, and showed that 
these were sufficient to evaluate the free energy of the model by Wiener-Hopf factorization. 
The first assumption
is very plausible and perfectly typical of the  inversion relation method, which needs 
such assumptions to complete it.
The second is by no means so obvious, implying as it does the symmetry (\ref{tau2symm}).
We have presented an argument for it - namely that $M_0$ is {\em not} a branch cut in 
the $t_q$ plane for the function $\tau_N(t_q) = \tau_2(t_q) \cdots \tau_2(\omega^{N-2} tq)$, and 
the easiest way to ensure this is to require that  $\tau_2(t_q)$ only have the cut $M_{N-1}$.
However, we must admit it is unlikely that we should have made such an assumption if we had 
not already known that it was implied by the exact finite-size functional relation method used in 
\cite{Seoul90}  and \cite{Baxter90}.

In Appendix B we present for reference the solution for the $N=2$ Ising case, where 
the standard elliptic 
function methods work immediately. However, in this case Assumption 2 is redundant, 
being a direct consequence of  $\tau_N(t_q)$ not having $M_0$ as a branch cut.

In later papers we intend to obtain the full Riemann surface, and all poles and zeros,  
of $T(x_p,y_p |  x_q,y_q)$ from (\ref{Trcont}), (\ref{2ndratio}) and (\ref{Trcontp}). This gives a very full description of 
the function and should greatly extend our understanding of it.

\section*{Appendix A: checking the equations}

\renewcommand{\theequation}{A\arabic{equation}}
\appendix
\setcounter{equation}{0}
When performing the intricate calculations necessary to derive equations such 
as (\ref{tau2result}), (\ref{Sresult}), one encounters
many irritating and uninteresting factors that are independent of $q$. Two checks
on these can be obtained from the invariances discussed near the end of section 2: these can be 
used to immediately check that the powers of $k$ and $y_p$ are correct. 
In particular,  $\tau_2(t_q)$ is proportional to $y_p^{-2}$, while
$\xi_q , S(\lambda_q)$ are proportional to $y_p^{-N(N-1)}$.
The functions $T(x_q,y_q)$ and $T(y_q,x_q)$ do not contain any such external $y_p$ factors.

We are still left with constant factors coming from products of powers of $\pm \omega^j$ and 
$1 - \omega^j$. 
To check these, the author has found it helpful to 
consider the low-temperature limit in which $k' \rightarrow 0$, $\lambda_p$ and 
$\lambda_q$ are finite, and $x_p, y_p, x_q,y_q  \rightarrow k^{-1/N}$, 
$t_p,  t_q  \rightarrow k^{-2/N}$. One needs to 
first eliminate zero differences like $y_q - x_p$, $t_p - t_q$ by using
exact formulae such as 
\be y_q^N - x_p^N \eq \frac{k'( 1 - \lambda_p \lambda_q )}{k \lambda_p} \sep t_p^N-t_q^N \eq 
\frac{k' (\lambda_p - \lambda_q)(1-\lambda_p \lambda_q)}{k^2 \lambda_p \lambda_q} \period \ee

One can also use formulae such as
\bd
\prod_{j=1}^{N-1} (1-\omega^j) = N \sep \prod_{j=1}^{N-1} (1-\omega^j)^j = N^{N/2} \, 
e^{i \pi (N-2)(N-1)/12}  \ed 
(though we have tried to arrange the expressions so that only the first of these is needed).

In this limit $T(x_q,y_q) = T(y_q,x_q) = 1$ and 
\be  \tau_2 (\omega^{j-1} t_q) \eq (1 - \omega^{j} )/(k^{2/N} y_p^2) \; \; \; \; {\rm for} \; \; j = 1, \ldots ,N-1 \comma \ee
while from (\ref{tauprod}) and (\ref{defalpha}),
\be \tau_2 (\omega^{-1} t_q) \eq \frac{k' (1-\lambda_p \lambda_q )^2}{N k^{2/N} y_p^2 \lambda_q} 
\period \ee

It follows  from (\ref{defxi}) that 
\be \xi_q \eq  \left\{ \frac{k' (1- \lambda_p \lambda_q) }
{ k \, y_p^N \lambda_p^{1/2} } \right\}^{N-1}  \ee
and from (\ref{defS2}) that
\be \label{Slowtemp}  S(\lambda_q) \eq \xi_q (1-\lambda_p \lambda_q)^{N-1} \eq
  \left\{ \frac{k' (1- \lambda_p \lambda_q)^2 }
{ k \, y_p^N \lambda_p^{1/2} } \right\}^{N-1}  \ee
The rhs of (\ref{Sl1l}) is now
\bd  \left\{ \frac{k' (\lambda_p - \lambda_q)(1-\lambda_p \lambda_q)}
{k \, y_p^N \lambda_p^{1/2} \lambda_q }  \right\}^{2N-2} \comma \ed
and we see that  (\ref{Sl1l}) is indeed satisfied in this low-temperature limit.

Also, from (\ref{defAB}),
\be 
A_{pq} =  B_{pq} = 0 \comma \ee
so (\ref{Slowtemp}) agrees with (\ref{Sresult}).

From (\ref{Dbar}) and (\ref{Pprod}), 
\be
\overline{D}_{pq} = 1 \sep P_{pq} = (\lambda_p/\lambda_q)^{(N-1)/2} \comma \ee
so (\ref{TpowerN}) is satisfied.

\section*{Appendix B: The Ising case}

\renewcommand{\theequation}{B\arabic{equation}}
\appendix
\setcounter{equation}{0}

When $N=2$, as in \cite{Baxter98a}, the parameters  $a_q, b_q, c_q, d_q$ of (\cite{BPAY88})
can be written as
\be a_q, b_q, c_q, d_q \eq -H(u_q), -H_1(u_q) , \Theta_1(u_q) , \Theta(u_q) \comma \ee
from which it follows that $x_q, y_q, \mu_q$ are
\be x_q = -\sqrt{k} \; {\rm sn} \, u_q \sep y_q = - 
\sqrt{k} \, \frac{{\rm cn} \, u_q}{{\rm dn} \, u_q} \sep \mu_q = 
\frac{ \sqrt{k'} }{ {\rm dn} \, u_q }\period \ee

Here $H(u), H_1(u), \Theta(u), \Theta_1(u), {\rm sn \,}u_q, {\rm cn \,}u_q, {\rm dn \,}u_q$ are the usual Jacobi elliptic functions 
of modulus $k$. For $(x_p,y_p), (x_q, (y_q) \in
 {\cal D}_0$ we can restrict $u_p$ and $u_q$ to lie in the corresponding rectangle in 
\uqplane, i.e. $ 3K/2 < \Re (u_q) < $ $ 5K/2$, $-K' < \Im (u_q) \leq K'$. (The variables and 
functions $x_q, y_q, t_q, \lambda_q,$ $ T(x_q,y_q)$, $\tau_2(t_q), S(\lambda_q)$ 
in this Appendix are all periodic of period $2 i K'$.)
Similarly, $(x_q,y_q) \in {\cal D}'_0$ (or ${\cal D}'_1$) if  $u_q$ lies in the rectangle 
${\cal D}'_0$ (or ${\cal D}'_1$) in  \uqplane. The vertical lines in between the rectangles 
are where $|\lambda_q| = 1$.

Defining
\bd {\rm scd}\, u = \frac{{\rm sn \,}u}{{\rm cn \,}u \; {\rm dn \,}u } 
 \sep h_1(u) = H_1(u/2) \Theta_1(u/2) \comma \ed
it follows that
\be
W_{pq}(1) = k' {\rm scd}\, \left(\frac{K-u}{2} \right) \sep \Wb_{pq}(1) = k' {\rm scd}\, 
\left(\frac{u}{2} \right) \comma
\ee
where $K$ ($K'$) is the complete elliptic integral of modulus $k$ ($k'$) and 
\be u = u_q - u_p  \ee
and by definition $W_{pq}(0) = \Wb_{pq}(0) = 1$.

\begin{figure}[hbt]
\begin{picture}(420,234) (-40,-34)
\put(15,25) {\line (1,0) {310}}
\put(15,175) {\line (1,0) {310}}
\thicklines
\put(35,10) {\line (0,1) {180}}
\thinlines
\put(70,15) {\line (0,1) {170}}
\put(140,15) {\line (0,1) {170}}
\put(210,15) {\line (0,1) {170}}
\put(280,15) {\line (0,1) {170}}
\put(23,87) {O}
\put(16,163) {$iK'$}
\put(7,13) {$-iK'$}
\put(32,97) {$\bullet$}
\put(67,97) {$\bullet$}
\put(137,97) {$\bullet$}
\put(207,97) {$\bullet$}
\put(277,97) {$\bullet$}
\put(73,87) {$K/2$}
\put(143,87) {$3K/2$}
\put(213,87) {$5K/2$}
\put(283,87) {$7K/2$}
\put(172,115) {${\cal D}_0$}
\put(242,115) {${\cal D}'_0$}
\put(102,115) {${\cal D}'_1$}
\put(10,-15) {Figure 4: The complex $u_q$ plane in the Ising case, showing the }
\put(50,-32){rectangles corresponding to ${\cal D}_0, {\cal D}'_0, {\cal D}'_{1}$.}
\end{picture}
\end{figure}

We note that the Boltzmann weights depend on $u_p$, $u_q$ only via their difference 
$u_q - u_p$. This is the ``difference property'' for this model (it {\em only} holds for 
$N = 2$). The partition function, transfer matrix  and hence $T(x_q,y_q)$ therefore also 
depend on $p$ and $q$ only via $u = u_q - u_p$; in particular we can write $T(x_q,y_q)$
as $T(u)$.

From (\ref{defrpq}),
\be \label{ru}
r_{pq} \eq r(u) \eq \frac{h_1(0) \Theta_1(0) H_1(u) }{h_1(u)^2 } \period \ee

Incrementing $u_q$ by $K$  takes $x_q, y_q$ to $y_q, -x_q$. Hence the inversion relation
(\ref{3.46_1}) becomes \be \label{invIsing}
T(u) T(u+K)  =  r(u) \period \ee

Define \be z = e^{- \pi u/K'} \sep q' = e^{-\pi K/K'} \period \ee
Our assumption 1 implies that $\log T(u)$ is analytic in the vertical strip
$0 \leq \Re  (u) \leq K$ in the complex $u$-plane, and periodic of period $2i K'$. Hence 
in this strip it has a Fourier expansion of the form
\be \log T(u) \eq \sum_{m= -\infty}^{\infty} \rho_m \, z^m \period \ee
Taking logs of (\ref{invIsing}), using this expansion and equating Fourier coefficients, we obtain
\be \label{Onsager} \log T(u) \eq \sum_{m=1}^{\infty} \frac{ ({q'}^m -{q'}^{2m})(1-z^m)(1-{q'}^m/z^m) }
{m \, (1+{q'}^m)^2 \, (1+{q'}^{2m}) } \comma \ee
which can be shown to be Onsager's result \cite{Onsager44} for the free energy of the Ising model.
If we define
\be g(z) \eq \prod_{m=0}^{\infty} \frac{ ( 1- {q'}^{4m+1} z )^{4m+1} \; ( 1- {q'}^{4m+3} z )^{4m+4}}
{ ( 1- {q'}^{4m+2} z )^{4m+3} \; ( 1- {q'}^{4m+4} z )^{4m+4}} \comma \ee
then we can write (\ref{Onsager}) as
\be \label{Tgggg}
T(u) \eq g(z) \, g(q'/z)/[ g(1) g(q') ] \comma \ee
which manifests the rotation symmetry (\ref{krotn}), i.e. $T(u) = T(K-u)$.

The auxiliary functions  $\tau_2(t_q)$, $S(\lambda_q)$ simplify if we define
\be \tilde{T} (u) =  h_1(u) \,
 h_1(K-u) \,
T(u) /[h_1(0) h_1(K)]\comma \ee
then
\be
\tau_2(t_q) \eq \frac{2 \,  h_1(K)^2 \, \tilde{T}(u_q-u_p) \, \tilde{T}(5K - u_q - u_p)  }
{ H_1(u_p)^2 \,
\Theta(u_q) \, \Theta_1(u_q) } \comma
 \ee
\be S(\lambda_q ) \eq \frac{4 \, h_1(K)^2 \, \tilde{T} (u_q-u_p) \, \tilde{T} (4K -u_q-u_p)}
{\left[ H_1(u_p) \Theta(u_q) \right]^2} \period \ee
The first of these  formulae explicitly manifests the invariance of $\tau_2(t_q)$ under 
$u_q \rightarrow 5K - u_q$, corresponding 
to interchanging   ${\cal D}_0$ with ${\cal D}'_0$ and $x_q$ with $y_q$. The second manifests the invariance of 
$S(\lambda_q)$ under $u_q \rightarrow 4K - u_q$, corresponding 
to  negating $x_q, t_q$  in ${\cal D}_0$, while leaving $\mu_q$, $\lambda_q$, $y_q$   
unchanged. Note that
\bd h_1(K)^2 = H_1(0)^2 \Theta(0) \Theta_1(0)/2 \period \ed

There are further simplifications from using $\tilde{T}(u)$ instead of $T(u)$.  The 
original inversion and rotation relations (\ref{inv1}), (\ref{krotn}) become
\be \tilde{T}(u) \tilde{T}(-u) = \frac{H_1(u)^2}{H_1(0)^2} \sep 
\tilde{T}(u) = \tilde{T}(K-u) \comma \ee
while (\ref{Tgggg}) becomes
\be\tilde{T}(u) \eq e^{\pi u(K-u)/4KK'} \, \tilde{g}(z) \, \tilde{g}(q'/z)/ \left[
\tilde{g}(1) \, \tilde{g}(q') \right]  \comma \ee
where $\tilde{g}(z)$ is a considerably simpler function than $g(z)$, being
\be
\tilde{g}(z)  = \prod_{n=1}^{\infty} \left( \frac{1-{q'}^{2n-1} z }
{1 - {q'}^{2n} z } \right)^{2n} \period \ee

\noindent It is also true that   
\be
\tilde{T}(u) = \frac{H_1(u)}{H_1(0)} \, e^{\pi u/4 K'} \, \prod_{n=1}^{\infty}
\frac{(1-{q'}^{2n-1} z )^{2n-1} (1-{q'}^{2n} z^{-1} )^{2n} }
{(1-{q'}^{2n-1} z^{-1} )^{2n-1} (1-{q'}^{2n} z )^{2n} } \period \ee



\begin{thebibliography}{99999}

\bibitem{Onsager44} L. Onsager, Crystal statistics. I. A two-dimensional model 
with an order-disorder transition, Phys. Rev 65 (1944) 117 -- 149.

\bibitem{Kaufman49} B. Kaufman,  Crystal statistics. II. Partition 
function evaluated by spinor analysis, Phys. Rev. 76 (1949) 1232 -- 1243.

\bibitem{Kasteleyn63} P.W. Kasteleyn, Dimer statistics and phase transitions, J. Math. 
Phys.  4 (1963) 287 -- 293.

\bibitem{Stroganov79} Yu.G. Stroganov, A new calculation method for partition functions 
in some lattice models, Phys. Lett. A 74 (1979) 116 -- 118. 

\bibitem{BaxterEnting78} R.J. Baxter and I.G. Enting, 399th solution of the 
Ising model, J. Phys. A 11 (1978) 2463 --2473.

\bibitem{Baxter88} R.J. Baxter, Free energy of the solvable chiral Potts 
model, J. Stat. Phys. 52 (1988) 639 -- 667.

\bibitem{Seoul90} R.J. Baxter, Calculation of the eigenvalues of of the transfer matrix
of the chiral Potts model, Proc. Fourth Asia-Pacific Physics Conference 
(Seoul, Korea, 1990), World-Scientific, Singapore (1),  1991,  42 -- 58.

\bibitem{Baxter90} R.J. Baxter,  Chiral Potts model: eigenvalues of the tranfer 
matrix, Phys. Lett. A  146 (1990) 110 -- 114.

\bibitem{Baxter00} R.J. Baxter, Equivalence of the two results for the free energy
of the chiral Potts model, J. Stat. Phys. 98 (2000) 513 -- 535.

\bibitem{Baxter91} R.J. Baxter, Hyperelliptic function parametrization for 
the chiral Potts model, Proc. Int. Congress Mathematicians (Kyoto, Japan, 1990),
 Springer-Verlag, Tokyo, 1991, 1305 -- 1317.

\bibitem{Baxter81} R.J. Baxter, Rogers-Ramanujan identities in the hard hexagon 
model, J. Stat. Phys.  26 (1981) 427 -- 452.

\bibitem{Howes83} S. Howes, L.P. Kadanoff and M. den Nijs, Quantum model for 
commensurate-incommensurate transitions, Nucl. 
Phys. B 215[FS7] (1983) 169 -- 208.

\bibitem{Albertini89} G. Albertini, B.M. McCoy, J.H.H. Perk and S. Tang,
Excitation spectrum and order parameter for the integrable $N$-state chiral 
Potts model, Nucl. Phys. B  314 (1989) 741 -- 763.

\bibitem{HenkelLacki89} M. Henkel and J. Lacki, Integrable chiral $Z_n$ quantum chains 
and a new class of trigonometric sums,  Phys. Lett. A 
138 (1989) 105 -- 109.

\bibitem{Baxter98a}  R.J. Baxter, Functional relations for the order parameters
of the chiral Potts model, J. Stat. Phys.  91 (1998) 499 -- 524.

\bibitem{Baxter98b}  R.J. Baxter, Some hyperelliptic function identities that 
occur in the chiral Potts model, J. Phys. A  31 (1998) 6807 -- 6818.

\bibitem{Baxter98c}  R.J. Baxter, Functional relations for the order parameters
of the chiral Potts model: low temperature expansions, 
Physica A 260 (1998) 117 -- 130.

\bibitem{BPAY88} R.J. Baxter, J.H.H. Perk and H. Au-Yang,  New solutions 
of the star-triangle relations for the chiral Potts model, Phys. Lett. A 
128 (1988) 138 -- 142.

\bibitem{BBP90} R.J. Baxter, V.V. Bazhanov and J.H.H. Perk, 
Functional relations for transfer matrices of the chiral Potts model,
Int. J. Mod. Phys. B 4 (1990) 803 -- 870.

\bibitem{BazStrog90} V.V. Bazhanov and Yu.G. Stroganov, Chiral Potts 
model as a descendant of the six-vertex model, J. Stat. Phys. 
59 (1990) 799 -- 817.

\bibitem{BaxSkew93} R.J. Baxter, Chiral Potts model with skewed boundary 
conditions, J. Stat. Phys. 73 (1993) 461 -- 495.

\bibitem{Baxter80} R.J. Baxter, Exactly Solved Models, {\it in}
Fundamental Problems in Statistical Mechanics V (ed. E.G.D. Cohen),
 North-Holland, Amsterdam, 1980, 109 -- 141.

\bibitem{Baxter82} R.J. Baxter, The inversion relation method for 
some two-dimensional exactly solved models in lattice statistics,
J. Stat. Phys. 28 (1982) 1 -- 41.

\bibitem {book82}  R.J. Baxter,
Exactly Solved Models in Statistical Mechanics,  Academic, London, 1982.

\bibitem{Noble58} B. Noble, Methods Based on the Wiener-Hopf Technique,
Pergamon, London, 1958.

\end{thebibliography}
\end{document}